\documentclass[12pt,a4paper]{article}
\usepackage{jheppub_kim}
\usepackage{epsfig}
\usepackage{amsmath}
\usepackage{graphicx}
\usepackage{graphics}
\usepackage[scriptsize,nooneline,hang]{caption}
\usepackage[hang,nooneline,scriptsize]{subfigure}

\begin{document}

\title{Spacetime expansion in the presence of a background velocity field}

\author{Solmaz Asgari, }

\author[*]{Reza Saffari}
\affiliation{Department of Physics, University of Guilan,
41335-1914, Rasht, Iran}

\emailAdd{rsk@guilan.ac.ir}

\abstract{In this article, we introduce a new metric assuming an additional velocity-based term in a spacetime metric. Although the inclusion of this additional phrase can indicate that the Lorentz symmetry has broken, the results of null geodesics demonstrate that the amount of variation in the speed of light is considerably smaller than what can be observed. This article delves into the primary use of this additional phrase in the expansion development of Minkowski and de Sitter spacetimes. The findings indicate that multiple versions of the velocity function may explain the universe's initial explosion and expansion behavior even before the inflation epoch, the inflation itself, and the universe's positive acceleration in late time cosmology without dark matter or dark energy. These models can also rebuild primordial black holes in the early universe as an alternative to dark matter.}

\maketitle

\section{Introduction}\label{intro}

Despite the passage of time since discovering the universe's expansion, many cosmological issues remain unanswered. Some of the questions that need to be addressed are the source of this growth and the dynamics. As we know, the cosmos is growing at a positive acceleration now \cite{Riess1998}. However, we are unsure precisely what is causing this rapid growth. The idea of dark energy is the most straightforward approach to solve the conundrum of accelerating expansion. Because of the negative pressure, dark energy is a unique type of energy. The cosmological constant is the most basic and suitable type of dark energy \cite{Peebles2003}. The current cosmological standard model, the $\Lambda$CDM model, includes a small positive cosmological constant in the Einstein field equations to describe the positive acceleration \cite{Carroll2001}, \cite{Steinhardt2006}. According to this hypothesis, today, the universe is made up of visible stuff, dark matter, and dark energy \cite{Planck2018}. It is also evident that the world expanded at a near-infinite rate for a brief period called inflation \cite{Starobinsky1980}-\cite{Linde1983}. It describes how quantum fluctuations in the tiny inflationary zone, as the seeds for the structure formation in the universe amplified to cosmic scale. Explaining how the universe seems to be isotropic, how cosmic microwave background radiation is equally distributed, why even the cosmos is flat, and yet no magnetic monopoles have been observed are only a few of its essential aspects \cite{Hawking1982}-\cite{Bardeen1983}. 

On the other hand, modified gravity is a theory that involves altering Einstein's field equation. These class of gravity theories can be categorized in a set of alternatives to general relativity. Following general relativity, efforts were undertaken to improve ideas created before general relativity or enhance general relativity itself. Many alternative techniques have been tried, such as adding spin to general relativity, combining a general relativity-like metric with a static spacetime concerning the universe's expansion, and adding another parameter to gain more flexibility. At least one theory was inspired by a desire to find a singularity-free alternative to general relativity. The ideas improved, as did the experimental testing. Unfortunately, many of the numerous techniques created shortly after general relativity were abandoned. Any extension of general relativity can substitute for this sort of theory, such as scalar field, quasilinear, tensor, scalar-tensor, vector-tensor, bimetric, and non-metric theories. \cite{Ni1972}-\cite{Rastall1979}. Instead, there was a rush to build more generic forms of the theories that remained so that a theory would be available if any test revealed a contradiction with general relativity. The motivations behind the more recent alternative hypotheses are nearly all cosmological, with concepts like inflation, dark matter, and dark energy being connected with or replacing them \cite{Bekenstein2004}-\cite{Lombriser2017}. As can be seen, despite the standard cosmological model's effectiveness in justifying observations of the cosmos, alternative gravity approaches are still being investigated as a means of overcoming the universe's dark content. 

When considering any expansion for the cosmos, we must eventually face some velocity functions, which are generally calculated after defining the spacetime's metric components. However, in this article, we will look at the velocity field as a separate portion of spacetime associated with a spacetime metric, affecting the universe's dynamics. We utilize Planck's time multiplied by velocity, which is the consequence of length, in this work, and it may thus be used with other metric sentences. Because the velocity function is unknown, we examine the simplest feasible model for it to measure just the ability to add this function to the metric as a preliminary study of the outcome of its existence in the metric.

The existence of a field of velocity in a metric means that every two points in spacetime, which have a spatial difference and a temporal difference, inevitably have a velocity difference as well. In Minkowski spacetime, every two points in spacetime have a spatial and a temporal difference, so the constant parameter is velocity. The presence of a constant quantity of velocity makes the Lorentz transformations in this spacetime constant. However, in the metric introduced in this paper, because there can be a velocity between every two spacetime points, the velocity parameter is no longer introduced as an invariant property of this spacetime. Therefore the constant acceleration parameter plays a pivotal role. 

This paper considers spacetime's evolution in the presence of a background velocity from the beginning of the universe as its purpose. The outline of this paper is as follows. In Sec. \ref{AE}, with the help of Hubble law, we assume a generic bivariate expansion velocity as a function of distance and time. In Sec. \ref{VDS}, we apply the background velocity for Minkowski spacetime, and we obtain the scalar curvature and null geodesics in the presence of the velocity field. In Sec. \ref{EdSS}, we apply the background velocity for de Sitter spacetime, and we consider the metric horizons in general cases. In Sec. \ref{TM}, we choose a toy model for the background velocity field with two degrees of freedom. Then we discuss the existence of primordial black holes, the big bang and inflation, big crunch, and late time accelerating expansion of the universe for different values of these two free parameters, and we finally conclude in Sec. \ref{C}.

\section{Accelerating expansion}\label{AE}

Edwin Hubble discovered the universe's expansion rate, which can be measured by the velocity at which a distant galaxy is moving away, as a function of the galaxy's distance from us \cite{Hubble1929}. This function's initial form represented a linear relationship between the velocity and the target galaxy's distance
\begin{equation}\label{AE1}
v = H x,
\end{equation}
where $H$ was called the Hubble constant. Today $H$ is called the Hubble parameter due to the acceleration in expansion. Then the Hubble parameter may vary from time to time in the history of the universe.

In the standard cosmological model, assuming the universe's homogeneity and isotropy principles, the Friedmann-Lemaitre-Robertson-Walker (FLRW) metric is chosen to describe this expansionary behavior. Using the FLRW metric, the definition of velocity transforms from distance to time
\begin{equation}\label{AE2}
x = R(t) X,
\end{equation}
where $R(t)$ is the dimensionless scale factor in which $t$ denotes the time from the birth of the universe, and $X$ is the coordinate distance. The time-dependent Hubble parameter,
\begin{equation}\label{AE21}
H(t)=\frac{\dot{R}(t)}{R(t)},
\end{equation}
where $\dot{R}$ is the time-derivative of the scale factor, describes the universe's expansion rate.

Now, the question is, if we assume the expansion rate of the universe not only depends on the distance, but also depends on time, how can we analyze the expansion velocity in a general form? So, in response of this question one can consider with the help of Hubble's observation in Eq. (\ref{AE1}), and its development, the velocity of galaxies can be assumed as a function of distance and a function of time; we define the bivariate velocity function as
\begin{equation}\label{AE3}
v \equiv v(x,t),
\end{equation}
then its total variation can obtain as
\begin{equation}\label{AE4}
dv = \frac{\partial v}{\partial x} dx + \frac{\partial v}{\partial t} dt.
\end{equation}
As a generic form of a bivariate function, one can write the velocity's Taylor expansion up to the first order of variables around a particular point of space-time $(x_0,t_0)$, as
\begin{equation}\label{AE5}
v(x,t)=v(x_0,t_0)+\frac{\partial v}{\partial x}\Big|_{(x_0,t_0)}(x-x_0)
+\frac{\partial v}{\partial t}\Big|_{(x_0,t_0)}(t-t_0),
\end{equation}
and its derivative gives the local differential of velocity as
\begin{equation}\label{AE6}
dv\Big|_{(x_0,t_0)} = \frac{\partial v}{\partial x}\Big|_{(x_0,t_0)} dx + \frac{\partial v}{\partial t}\Big|_{(x_0,t_0)} dt.
\end{equation}

The Eqs. (\ref{AE4}) and (\ref{AE6}) are the global and the local forms of the velocity evolution as a bivariate function. In each of the above equations, the term ${\partial v}/{\partial x}$, denotes the variation of velocity over distance which is known as Hubble constant (refers to Eq. (\ref{AE6})) or Hubble parameter (refers to Eq. (\ref{AE4})). Also, the term ${\partial v}/{\partial t}$ in both (\ref{AE4}) and (\ref{AE6}) equations, introduces the pure acceleration which can describe as a general and a locale variation of velocity over time, respectively. So, by comparison with Eq. (\ref{AE4}) one can write the evolution of velocity in its differential form as
\begin{equation}\label{AE7}
dv = H(x,t) dx + a(x,t) dt,
\end{equation}
where $H(x,t)\equiv \partial v/\partial x$ and $a(x,t)\equiv \partial v/\partial t$ are unknown general parameters. One can rewrite Eq. (\ref{AE6}) to describe the local differential evolution of velocity as
\begin{equation}\label{AE8}
dv\Big|_{(x_0,t_0)} = H_0 dx + a_0 dt.
\end{equation}
where $H_0$ and $a_0$ are the present values of measurable parameters. In this case, in an expanding universe, the basic premise is that the velocity rate relative to space can be independent of the rate of velocity change over time. Of course, this does not mean that these two parameters are never related during the expansion process, but we will see that they can be compatible in certain conditions. As it follows from Eq. (\ref{AE8}), assuming that both measurable parameters are positive, if we ignore one of the statements, the velocity rate will decrease, precisely when we need supposing an extra component such as dark energy to compensate for this shortcoming.

\section{Minkowski sacetime in the presence of a velocity depended background}\label{VDS}

A flat spacetime or a special relativistic metric in spherical coordinates is
\begin{equation}\label{CC4}
ds^2= - c^2 dt^2 + dx^2 + x^2 d\Omega^2,
\end{equation}
where $x$ is the radial coordinate and $d\Omega^2=(d\theta^2 + \sin^2\theta d\phi^2)$. If we assume that a field of velocity causes the universe's expansion, then we can consider this velocity field between every two points in spacetime. Therefore, to generalize the metric, we add a term in length dimension combined from a constant time and a general velocity to the metric (\ref{CC4}) as
\begin{equation}\label{CC2p}
ds^2= - c^2 dt^2 + dx^2 + x^2 d\Omega^2 + \epsilon~t_p^2~dv^2,
\end{equation}
where $\epsilon$ is a dimensionless constant with three values $\epsilon = 0, \pm 1$. In the case of $\epsilon = 0$, the generalization will vanish. In the case of $\epsilon = +1, -1$, the additional term of the metric will have a spacelike and timelike manner, respectively. For simplicity, in this article, we assume no angular dependency for velocity term appeared in the metric, then the matrix representation of the new metric obtains as
\begin{equation}\label{Met1}
g_{\mu \nu}=\left(
\begin{array}{cccc}
 -c^2 + \epsilon t_p^2 (\frac{\partial v}{\partial t})^2 & \epsilon t_p^2 (\frac{\partial v}{\partial t}) (\frac{\partial v}{\partial x}) & 0 & 0 \\
 \epsilon  t_p^2 (\frac{\partial v}{\partial t}) (\frac{\partial v}{\partial x}) & 1+ \epsilon  t_p^2 (\frac{\partial v}{\partial x})^2 & 0 & 0 \\
 0 & 0 & x^2 & 0 \\
 0 & 0 & 0 & x^2 \sin ^2\theta \\
\end{array}
\right),
\end{equation}
then, the inverse metric obtains as
\begin{equation}\label{Met1inv}
g^{\mu \nu}=\left(
\begin{array}{cccc}
 -\frac1A[1 + \epsilon t_p^2  (\frac{\partial v}{\partial x})^2] & \frac1A[\epsilon t_p^2  (\frac{\partial v}{\partial t}) (\frac{\partial v}{\partial x})] & 0 & 0 \\
 \frac1A[\epsilon t_p^2  (\frac{\partial v}{\partial t}) (\frac{\partial v}{\partial x})] & \frac1A[c^2-\epsilon t_p^2  (\frac{\partial v}{\partial t})^2] & 0 & 0 \\
 0 & 0 & \frac{1}{x^2} & 0 \\
 0 & 0 & 0 & \frac{1}{x^2 \sin ^2\theta} \\
\end{array}
\right),
\end{equation}
where $A=c^2( 1 + \epsilon t_p^2  (\frac{\partial v}{\partial x})^2 )-\epsilon t_p^2  (\frac{\partial v}{\partial t})^2$. So, all the non-zero connections, Riemann tensor components, Ricci tensor components, and of course, Einstein tensor components are accessible. It also can be computed the covariant derivative of the new metric and can be verified that it is zero.

In this metric, the combination of $t_p v$, where $t_p$ is the Planck time, has created in contrast with $c t$, where $c$ is the velocity of light. The constant $c$ in $ct$ is replaced by a variable $v$ in $t_p v$, and the variable term, $t$ in $ct$, is changed to the constant $t_p$ in $t_p v$. The additional term in metric (\ref{CC2p}) means a velocity field between every two points in spacetime. Regardless of the amount of velocity, the coefficient $t_p\simeq 10^{-44}~s$ is so tiny that its effect on the deviation from flat spacetime can be negligible. It can also be pointed out that adding this sentence violates the Lorentz symmetry in coordinate transformations, but it is negligible in the presence of a small field of velocity around a local frame of reference. One can replace the general variation of velocity according to Eq. (\ref{AE7}) into Eq. (\ref{CC2p}) to obtain
\begin{equation}\label{CC2xtv}
ds^2 = - c^2 \Big(1-\epsilon\frac{t_p^2}{c^2}{a^2}\Big)dt^2+\Big(1+\epsilon t_p^2{H^2}\Big)dx^2+ 2\epsilon t_p^2 a H dt dx +x^2 d\Omega^2.
\end{equation}
The given metric appears to be singular on two physically relevant horizons. When the metric's purely temporal component, $g_{tt}$, changes signs from positive to negative, one apparent singularity occurs. Solving the equation $g_{tt} = 0$ for the case $\epsilon=1$ produces $a=\pm a_p=\pm c/t_p$, also known as Planck acceleration. The case $a=a_p\simeq 10^{52}~m/s^{2}$ shows a huge expansion acceleration in the evolution of the horizon, such as an inflationary acceleration of the early universe. The case $a=-a_p$ is another huge negative acceleration solution which can reduce an infinite velocity of expansion to a sensible value, such as an expansion rate observational cosmologists find after the inflation. The second horizon happens when the metric's solely radial component, $g_{xx}$, reaches infinity. $H\rightarrow\pm\infty$ is obtained by solving the quadratic equation $1/g_{xx} = 0$. The positive solution is in contraction to $a=a_p$, and the negative solution is compatible with $a=-a_p$. In the case of $\epsilon=-1$ the results are not real solutions.

\subsection{Scalar curvature}

The scalar curvature, often known as the Ricci scalar, is the essential curvature invariant of a Riemannian manifold. It gives each point on a Riemannian manifold one unique real number depending on the manifold's intrinsic geometry at that position.
The Ricci scalar of the metric (\ref{CC2p}) can rewrite in terms of $H$ and $a$ with the help of Eq. (\ref{AE4}) and Eq. (\ref{AE7})
\begin{eqnarray}\label{riccis2}
R &=& \frac{2 \epsilon l_p^2 }{t_p^4 x^2 [ \epsilon(a^2-c^2 H^2) -a_p^2]^2} [\epsilon l_p^2 H^4 +c^2 H^2 \left(1-\epsilon a^2/a_p^2 + 4\epsilon x \dot{H} a/a_p^2\right) \nonumber \\
&+& 2c^2 x H' H \left(1-\epsilon a^2/a_p^2\right) - x \dot{a} \left(2 \epsilon t_p^2 H^3+2 H+x H' \right)+x^2 \dot{H}^2]
\end{eqnarray}
where $l_p = c t_p$, is the Planck length, "~$\dot{}\equiv \partial /\partial t$", "$~'\equiv \partial/\partial x$", and $\frac{\partial^2 v}{\partial t \partial x}=\frac{\partial^2 v}{\partial x \partial t}=\dot{H}=a'$. The denominator of Eq. (\ref{riccis2}) is always positive. Then the sign of scalar curvature strongly depends on the sign of the numerator and the value of $\epsilon$. The roots of the denominator can obtain as $x=0$ or the solutions of $\epsilon(a^2-c^2H^2)-a_p^2=0$, which depend on the value of $\epsilon$. If $\epsilon=1$ then one solution obtains as $a^2=a_p^2+c^2H^2$, and if $\epsilon=-1$ then the other solution obtains as $a^2+a_p^2=c^2H^2$. Then $a$, $a_p$, and $cH$ behave such three sides of a right triangle in Pythagorean relation. It means the temporal part of the acceleration, $a$, is greater than the spatial part $cH$ when $\epsilon=1$, and vice versa. Both cases give infinite scalar curvature. Therefore, since the primitive universe requires an infinite amount of curvature, the metric introduced in Eq. (\ref{CC2p}) can supply that value in various forms, when the denominator of Eq. (\ref{riccis2}) is equivalent to zero, without the need for non-geometric content.

\subsection{Null geodesic}

A null geodesic is a path taken by a massless particle like a photon. It is named null since its interval is equal to zero, $ds^2=0$, and it does not have a proper time connected with it. In the case of a radial emission of light, $d\Omega^2=0$, the quadratic equation the of metric (\ref{CC2xtv}) which obtains from $ds^2=0$ is
\begin{equation}\label{ng1}
(1+\epsilon t_p^2 H^2)\Big(\frac{dx}{dt}\Big)^2 + 2\epsilon t_p^2 a H \Big(\frac{dx}{dt}\Big) - c^2(1-\epsilon \frac{t_p^2}{c^2}a^2)=0.
\end{equation}
The general solution obtains as
\begin{equation}\label{ng2}
\frac{dx}{dt}=\frac{-\epsilon t_p^2 a H \pm c \sqrt{1+\epsilon t_p^2 (H^2 - a^2/c^2)}}{1+\epsilon t_p^2 H^2}.
\end{equation}
Since $t_p^2 \ll 1$, then in local spacetime, incoming and outgoing light speeds are $-c$ and $+c$, respectively. One can also calculate the velocity of light in a special case $a^2=c^2H^2$. For the case $a=cH$, Eq. (\ref{ng2}) reduces to two separate solutions, such as
\begin{equation}\label{ng31}
\frac{dx}{dt}\Big|_{+}= \Big(\frac{1-\epsilon a^2/a_p^2}{1+\epsilon a^2/a_p^2}\Big)c,
\end{equation}
and
\begin{equation}\label{ng32}
\frac{dx}{dt}\Big|_{-}= -c.
\end{equation}
In the case $a=-cH$, Eq. (\ref{ng2}) reduces to other two different solutions
\begin{equation}\label{ng41}
\frac{dx}{dt}\Big|_{+}= c,
\end{equation}
and
\begin{equation}\label{ng42}
\frac{dx}{dt}\Big|_{-}= -\Big(\frac{1-\epsilon a^2/a_p^2}{1+\epsilon a^2/a_p^2}\Big)c.
\end{equation}
In all solutions from Eq. (\ref{ng31}) to Eq. (\ref{ng42}), the indices $+$ and $-$ on the left-hand side refers to the direction of light which can be outward and inward, respectively, and also different signs in Eq. (\ref{ng2}). Eqs. (\ref{ng32}) and (\ref{ng41}) show a familiar velocity of light for incoming and outgoing emitted lights, however Eqs. (\ref{ng31}) and (\ref{ng42}) seem to violate the constant velocity of light principle. In the case of $\epsilon=1$, the velocity of light obtains smaller than $c$, but for $a\ll a_p$, the deviation from $c$ cannot be detected because the additional term $a/a_p$ is still relatively smaller than one around the observer. In the case of $\epsilon=-1$, light speed obtains smaller than $c$ in both directions, but the deviations from $c$ is also negligible when $a\ll a_p$.

So far, we have come to conclude this section with the result that the new metric introduced in this article, in which a velocity field has been added to the Minkowski static spacetime, although not a problem in general relativity tensor calculations, is capable of generating an expanding spacetime at very high acceleration started at the beginning of the time. So, in following, we are interested to have a look on the capabilities of this velocity field in the presence of de Sitter spacetime.

\section{de Sitter spacetime in the presence of a velocity depended background}\label{EdSS}

The de Sitter spacetime is still the most symmetric vacuum solution of general relativity field equations with a positive cosmological constant. It is also well recognized for its use in Einstein's field solutions, which is one of the most fundamental mathematical models of the universe consistent with observed accelerated expansion. A four-dimensional de Sitter spacetime in spherical coordinates appears as
\begin{equation}\label{dS4}
ds^2= - c^2 (1 - \Lambda x^2) dt^2 + (1 - \Lambda x^2)^{-1} dx^2 +
x^2 d\Omega^2.
\end{equation}
Here we add the extra term of velocity as a perturbation term
\begin{equation}\label{dS2p}
ds^2= - c^2 (1 - \Lambda x^2) dt^2 + (1 - \Lambda x^2)^{-1} dx^2 +
x^2 d\Omega^2 + \epsilon t_p^2 dv^2.
\end{equation}
One can replace $dv$ from Eq. (\ref{AE7}) into Eq. (\ref{dS2p}) to obtain the general form of velocity depended on the de Sitter metric as
\begin{equation}\label{dS2xtv}
ds^2 = - c^2 \Big(1-\Lambda x^2-\epsilon\frac{t_p^2}{c^2}{a^2}\Big)dt^2+\Big(\frac{1}{1-\Lambda x^2} + \epsilon t_p^2{H^2}\Big)dx^2+ 2\epsilon t_p^2 a H dt dx + x^2 d\Omega^2.
\end{equation}
To find one of the horizons of this perturbed metric, we solve $g_{00}=0$. There are two solutions of the form
\begin{equation}\label{hds1}
x = \pm \frac{1}{\sqrt{\Lambda}}\Big(1-\epsilon \frac{a^2}{a_p^2}\Big)^{1/2},
\end{equation}
where the negative solution is physically unacceptable. So, in the case of $\epsilon=1$, if one assumes $a \rightarrow a_p$, then obtains $x \rightarrow 0$, which is following the inflationary universe's situation at early times. Here if one assumes $a \rightarrow 0$, then obtains $x \rightarrow 1/\sqrt{\Lambda}$, which is compatible with the late time radius of the universe in de Sitter solution. On the other hand, in the case of $\epsilon=-1$, if one assumes $a \rightarrow a_p$, then obtains $x \rightarrow \sqrt{2/\Lambda}$. However, if one assumes $a \rightarrow 0$, then obtains $x \rightarrow 1/\sqrt{\Lambda}$. Therefore, while the $\epsilon=1$ universe expands from zero to a specified $1/\sqrt{\Lambda}$ radius, the other $\epsilon=-1$ universe is contracting from radius $\sqrt{2/\Lambda}$ to radius $1/\sqrt{\Lambda}$.

On the other hand, we solve the quadratic equation
\begin{equation}\label{gxx}
\frac{1}{g_{xx}}=\frac{1-\Lambda x^2}{1+\epsilon t_p^2 H^2 (1-\Lambda x^2)}=0,
\end{equation}
which yields two solutions, $x=\pm1/\sqrt{\Lambda}$ for both $\epsilon=\pm 1$, which are by de Sitter solutions. For $\epsilon=1$, other solutions correspond to $H=\pm\infty$ which is the same as horizon's solutions in last section, and for $\epsilon=-1$, results are not real solutions.

Thus, we observed the general effects of the velocity field as a correction of various metrics. Now it is time to study the possible effects of certain models of this velocity field.

\section{A model of background velocity}\label{TM}

In the previous sections, the calculations were presented in their general forms. Both parameters, $H$ and $a$ were introduced in general, and we did not provide an example of how to evaluate them. In this section, we want to examine the results of choosing a function for $v(x,t)$, and therefore the two parameters $H$ and $a$ in the form of a particular example.
\begin{equation}\label{TM1}
v(x,t)=v_0 \Big(\frac{x}{x_0}\Big)^m \Big(\frac{t_0}{t}\Big)^n,
\end{equation}
where $v_0$, $x_0$, $t_0$, $m$, and $n$ are some arbitrary constants. However, the results obtained in this section can be the basis for selecting models that are more compatible with the evolution of the world. As an example we define

\subsection{Primordial Black Hole}
\begin{figure}[h]
\centering
\includegraphics{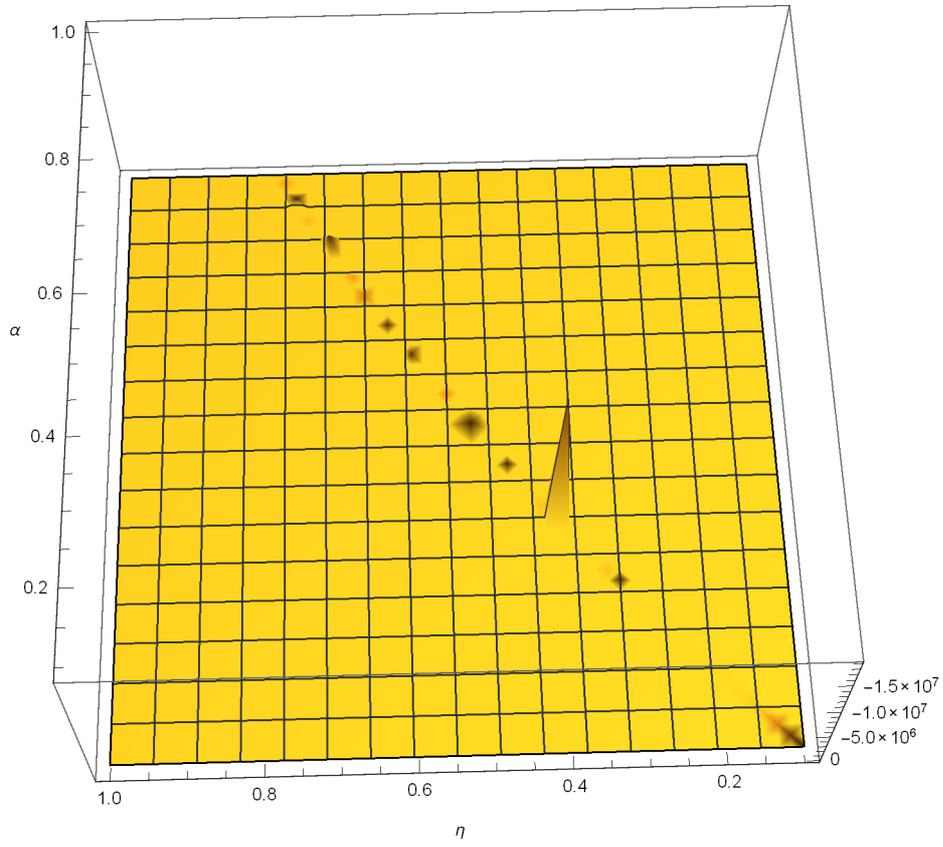}
\caption{\label{eta-alpha-3d-eps1} A three-dimensional plot of dimensionless scalar curvature in $\epsilon= +1$ illustrates primordial black holes in the early universe.}
\end{figure}

\begin{figure}[h]
\centering
\includegraphics{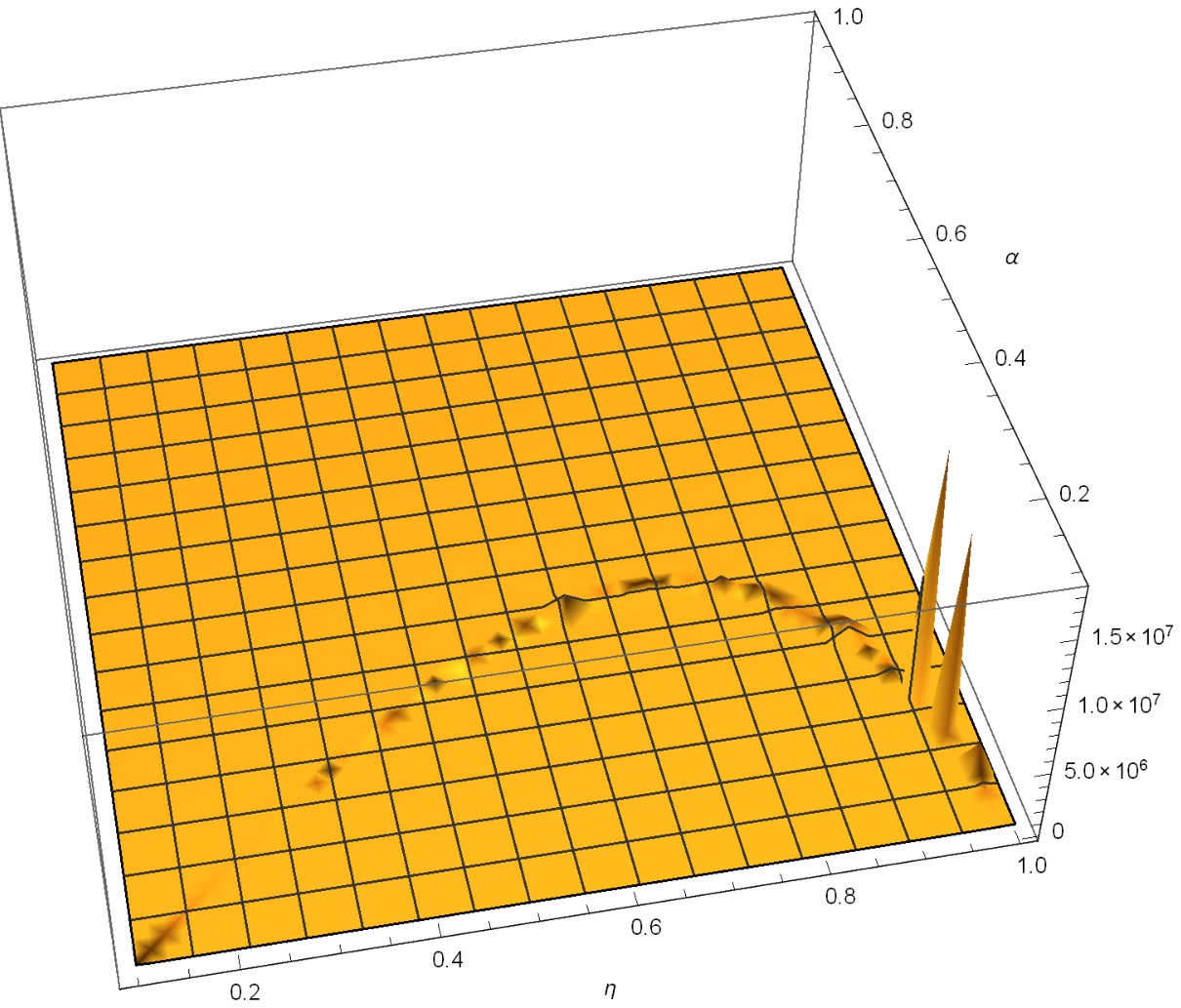}
\caption{\label{eta-alpha-3d-eps-1} A three-dimensional plot of dimensionless scalar curvature in $\epsilon= -1$ illustrates primordial black holes in the early universe.}
\end{figure}

\begin{figure}
\centering
\includegraphics{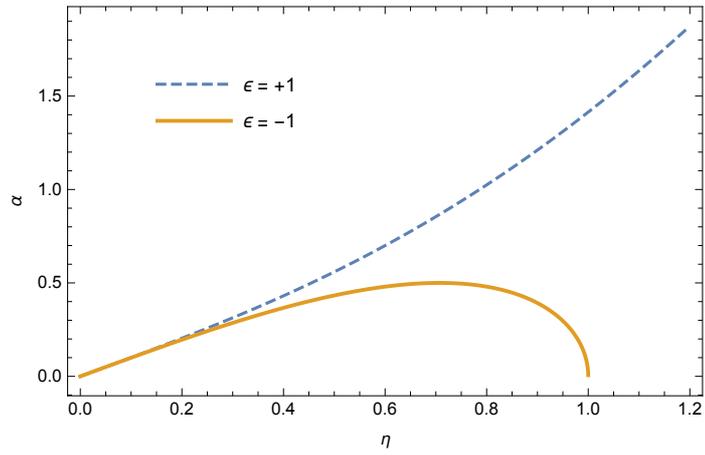}
\caption{\label{eta-alpha-2d-Rinf} The $\eta-\alpha$ relation for both cases of $\epsilon$ diverges the scalar curvature in the early universe.}
\end{figure}

In a simple case, when $v_0=c$, $x_0=l_p$, and $t_0=t_p$, metric in Eq. (\ref{CC2p}) takes the form
\begin{equation}\label{PBH}
ds^2= -c^2 {dt}^2 + {dx}^2 + \epsilon  \left(\frac{t_p}{t}\right)^{2 n} \left(\frac{x}{l_p}\right)^{2(m-1)} \left(m~t~dx - n~x\frac{dt}{t}\right)^2+ x^2d\Omega^2.
\end{equation}
Then the Ricci scalar obtains as appeared in Eq. (\ref{riccis2}). One can rewrite this equation in the form of a dimensionless function for $m=1$, and $n=1$ as
\begin{equation}\label{riccis3}
\frac{R}{R_p}=\frac{2 \epsilon \eta ^2 \left(\eta ^2 \left(\epsilon + \eta ^2 \right)-\alpha ^2 \left(\epsilon + 3 \eta ^2 \right)\right)}{\alpha ^2 \left(\epsilon \alpha ^2 -\eta ^2 \left(\epsilon + \eta ^2 \right)\right)^2},
\end{equation}
where, $R_p=1/l_p^2$, $t=\eta t_p$, $x=\alpha l_p$, and $l_p=c t_p$. Fig. \ref{eta-alpha-3d-eps1} and Fig. \ref{eta-alpha-3d-eps-1} show three-dimensional illustrations of the normalized scalar curvature for $\epsilon= \pm 1$, respectively. As shown in these figures, there are many local fluctuations in scalar curvature, which can be interpreted as primordial black holes in the early universe. The heights of fluctuations depend on the spacetime coordinates where the scalar curvature becomes divergent. The roots of the denominator in Eq. (\ref{riccis3}) determine these fluctuations. There are three solutions for
\begin{equation}
\alpha^2 (\epsilon \alpha^2 -\eta^2(\epsilon + \eta^2))^2=0,
\end{equation}
as
\begin{equation}\label{riccis4}
\alpha =0, \pm \eta  \sqrt{1 + \epsilon \eta ^2 },
\end{equation}
which diverges the Ricci scalar. These solutions also diverges the Kretschmann scalar of metric (\ref{CC2xtv}).
The case $\alpha=0$ causes the singularity of the primordial spacetime when the curvature of the universe is infinite. The other acceptable solution is the positive one which is the physical solution. As illustrated in Fig. \ref{eta-alpha-2d-Rinf}, along the $\eta-\alpha$ line, some curvature fluctuations could exist with different values of magnitude. In three-dimensional graphs, local curvature fluctuations have appeared.

\begin{figure}
\centering
\includegraphics{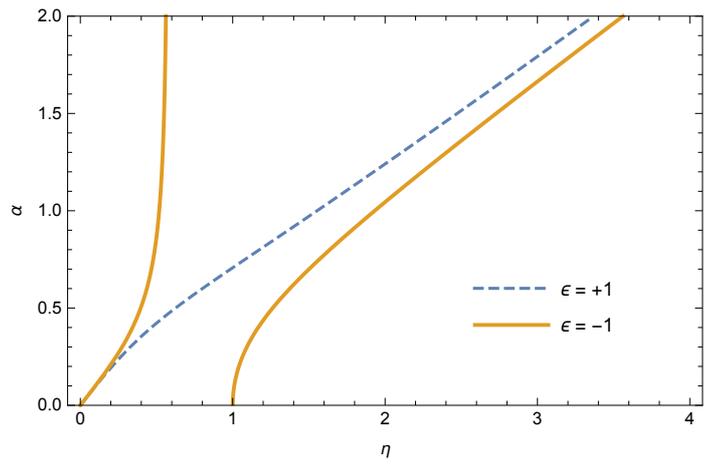}
\caption{\label{eta-alpha-2d-Rzero} The $\eta-\alpha$ relation disappears the scalar curvature and creates local flat spacetimes.}
\end{figure}

In the case of $\epsilon=1$, Fig. \ref{eta-alpha-3d-eps1}, curvature fluctuations move along the related $\eta-\alpha$ line, while in the case of $\epsilon=-1$, Fig. \ref{eta-alpha-3d-eps-1}, they will stop in a moment and move backward to reach the singularity $\alpha=0$. These phenomena may relate to the story of primordial black holes, which is in debates. Primordial black holes are a type of black hole that is thought to have formed shortly after the Big Bang. High densities and diverse circumstances in the early cosmos could have caused sufficiently dense places to collapse gravitationally, generating black holes. In 1966, Zel'dovich and Novikov hypothesized the existence of such black holes for the first time \cite{Zel1966}. Stephen Hawking was the first to investigate the theory of their genesis in-depth in 1971 \cite{Hawking1971}. They are a natural dark matter candidate since they are collision-free and stable, have non-relativistic velocities, and develop higher in the early universe. The same behavior of curvature scalar fluctuations happens for the metric (\ref{dS2xtv}). In this case, the denominator of the Ricci scalar is
\begin{equation}
\alpha^2 \left(\epsilon \alpha ^2 - \left(1-\Lambda l_p^2 \alpha^2 \right) \eta ^2 \left(\epsilon \left(1- \Lambda l_p^2 \alpha ^2\right)+ \eta^2 \right)\right)^2,
\end{equation}
in which the component $l_p^2 \Lambda \simeq 10^{-122}$ is negligible, so the $\eta-\alpha$ relation is the same as Eq. (\ref{riccis4}). It is also remarkable to find the $\eta-\alpha$ relation in which the Ricci scalar disappears. This relation obtains when one solves $\eta ^2 \left(\epsilon + \eta ^2\right)-\alpha^2 \left(\epsilon + 3\eta ^2\right)=0$ from the numerator of Eq. (\ref{riccis3}). This equation also has two solutions for $\alpha$ to $\eta$ as
\begin{equation}\label{riccis5}
\alpha = \pm \eta \sqrt{\frac{\epsilon + \eta ^2}{\epsilon + 3 \eta ^2}}.
\end{equation}
The positive solution in which the Ricci scalar disappears is desirable. The behavior of $\eta-\alpha$ relation is illustrated in Fig. \ref{eta-alpha-2d-Rzero} for both $\epsilon= \pm 1$. As it is shown in this figure, the expanding universe can have zero curvature for specific values of spacetime coordinates. The proportion that these $\eta-\alpha$ lines represent seems to be a subtle adjustment of how the universe expands while at the same time keeping its curvature somewhat close to zero.

\subsection{Big Bang and Inflation}
The Big Bang theory is the most broadly agreed cosmological model for explaining the visible universe's beginnings and rapidly growing evolution. The model describes how the universe developed from an elevated density and temperature initial points and gives a comprehensive description for a wide range of relevant aspects, such as light element abundances, the CMB, large-scale structure, and Hubble's law. On the other hand, cosmic inflation is the universe's idea growing exponentially in the initial cosmic eras. The inflationary period went through $10^{-36}$ seconds from the hypothesized Big Bang singularity to whenever it is between $10^{-33}$ and $10^{-32}$ seconds just after the singularity. The cosmos stretched to grow after the inflationary epoch, though at a lesser rate. To investigate the Big Bang and inflation in the presence of velocity background field introduced in Eq. (\ref{TM1}), one can also solve $g_{tt}=0$ from Eq. (\ref{dS2xtv}) to find the horizon's dynamics. Then the solution obtains as
\begin{equation}\label{TM2}
\epsilon n^2 \frac{\lambda^{2m}}{\tau^{2n+2}} + \lambda^2 - 1 =0,
\end{equation}
where $\lambda=x\sqrt{\Lambda}$, and $\tau=\lambda_p^{\frac{m}{2n+2}}t/t_p$ are two dimensionless parameters in which $\lambda_p=l_p \sqrt{\Lambda}$. Since Eq. (\ref{TM2}) cannot be solved for all values of $m$ at the same time, we solve it for only one value of $m$ at a time, and then we choose $m=1$ as the first solution. In this case, it has two solutions, such as
\begin{equation}\label{TM3}
\lambda(\tau)=\pm\frac{\tau^{n+1}}{\sqrt{\epsilon n^2 + \tau^{2n+2}}};
\end{equation}
one of them is physical solution because the negative solution is forbidden. Here we see if $\tau\rightarrow 0$ then $\lambda\rightarrow 0$ and if $\tau\rightarrow\infty$ then $\lambda\rightarrow 1$, which follows the early time radius and the late time visible radius of the universe in de Sitter spacetime. Notably, $\epsilon=-1$ is an unacceptable value because for $\tau<\sqrt[n+1]{n}$, the expression under radical will be negative. So, $\lambda$ appears as the expanding horizon's radius, $d\lambda/d\tau$ roles as expansion velocity, and $d^2\lambda/d\tau^2$ shows the horizon's acceleration. These parameters change with respect to normalized time $\tau$.

\begin{figure}
\centering
\includegraphics{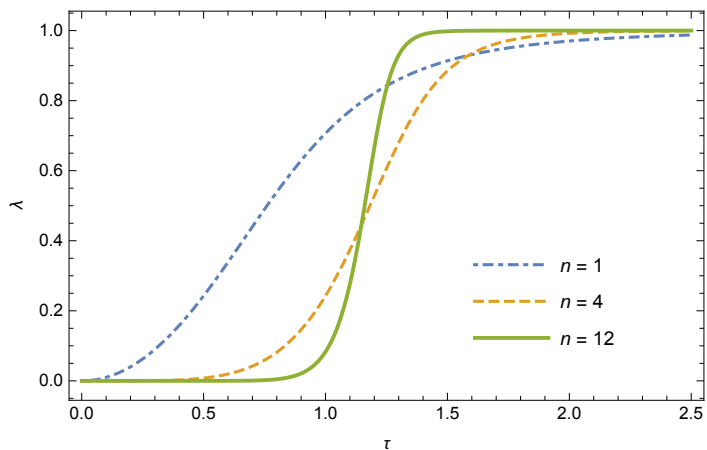}
\caption{\label{x-eps_1-m_1-n_1_4_12} The evolution of horizon's radius for three model-based spacetimes in the case of dimensionless parameters is depicted. In all of these models, $m=1$.}
\end{figure}

\begin{figure}
\centering
\includegraphics{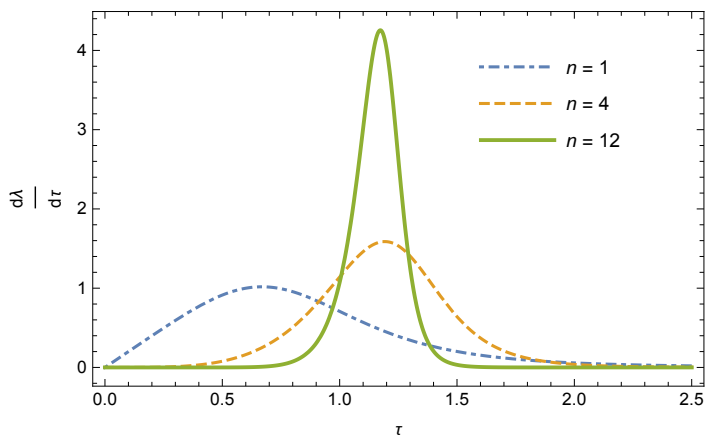}
\caption{\label{v-eps_1-m_1-n_1_4_12} The evolution of horizon's velocity for three model-based spacetimes in the case of dimensionless parameters is illustrated. In all of these models, $m=1$.}
\end{figure}

\begin{figure}
\centering
\includegraphics{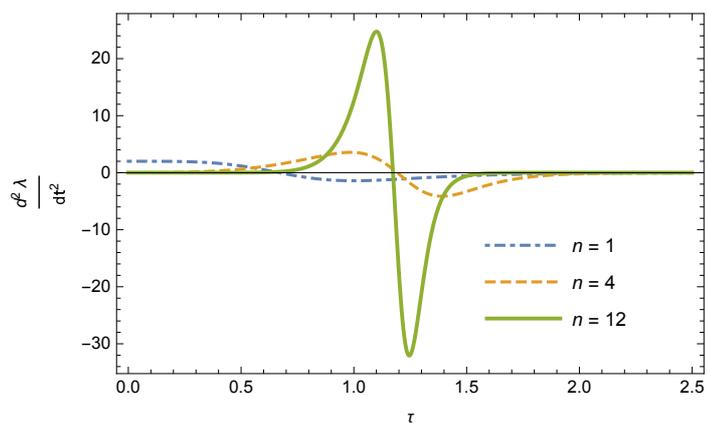}
\caption{\label{a-eps_1-m_1-n_1_4_12} The evolution of horizon's acceleration for three model-based spacetimes in the case of dimensionless parameters is shown. In all of these models, $m=1$.}
\end{figure}

As shown in Fig. \ref{x-eps_1-m_1-n_1_4_12}, the horizon's expansion has three stages in the presence of a background velocity field. The first step starts in $\tau=0$, and then the horizon expands gradually until the expansion rate increases sharply in the second stage. In the second expansion, the horizon's radius almost reaches the maximum value, and in the third stage, the expansion rate decreases slowly. These stages are very similar to the three stages of the universe's expansion from Big Bang to inflation, inflation by itself, and from inflation to late time expansion. This order of behaviors is typical for any option of values for parameter $n$, as illustrated in Fig. \ref{x-eps_1-m_1-n_1_4_12}. The only difference is the tangent of the graphs, which depends on the values of $n$. As it is appeared the inflationary period takes place in a shorter span when $n$ increases. For the cases of horizon velocity and horizon acceleration, the changing profiles between various steps and the contrasts between the values of $n$ appear in Fig. \ref{v-eps_1-m_1-n_1_4_12} and Fig. \ref{a-eps_1-m_1-n_1_4_12}, respectively.

\subsection{Big Crunch}
\begin{figure}
\centering
\includegraphics{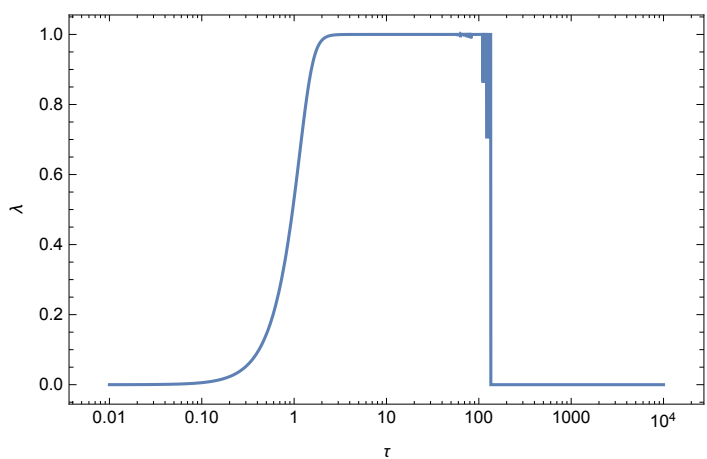}
\caption{\label{x-eps_1-m_2-n_3-t_-2_4} The evolution of an expanding universe's radius ends up in a crunch plotted over dimensionless time. In this model, $m=2$ and $n=3$.}
\end{figure}

\begin{figure}
\centering
\includegraphics{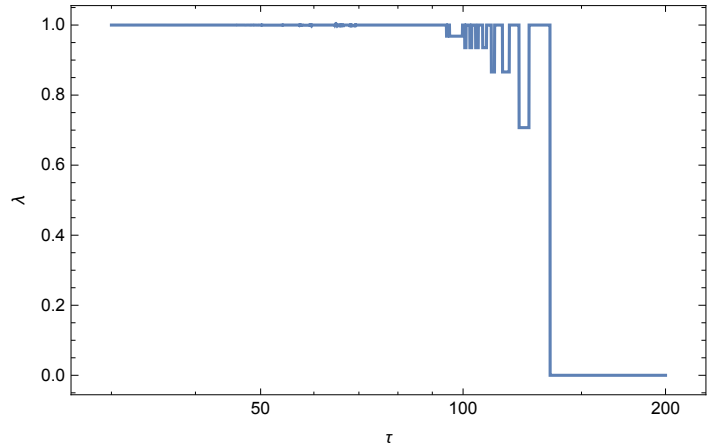}
\caption{\label{x-eps_1-m_2-n_3-t_30_200} A time scale of the model $m=2$ and $n=3$, illustrates radius fluctuations before the crunch is depicted.}
\end{figure}

The Big Crunch is a possible scenario for the universe's ultimate fate, in which the universe's expansion finally reverses and recollapses, causing the cosmic scale factor to approach zero, perhaps followed by a reformation of the cosmos beginning with another Big Bang. The overwhelming bulk of data contradicts this viewpoint. Rather than being restrained by gravity, astronomical data reveal that the expansion of the cosmos is speeding, implying that the cosmos is significantly more likely to terminate in heat death or a Big Rip. As another model of background velocity, one can choose $m=2$ in Eq. (\ref{TM1}). This case has four solutions as
\begin{equation}\label{TM4}
\lambda(\tau)=\pm\frac{\tau^{n+1}}{n}\sqrt{\frac{1}{2\epsilon}}\Big(\pm\sqrt{\frac{4 n^2 \epsilon}{\tau^{2n+2}}+1}-1\Big)^{1/2},
\end{equation}
that one of them is a physical solution. In this case, the evolution of the horizon is such as depicted in Fig. \ref{x-eps_1-m_2-n_3-t_-2_4}. In this case, the universe has a starting point like what is seen in the case of $m=1$, but it fails to zero radii after experiencing an inflationary epoch. As shown in Fig. \ref{x-eps_1-m_2-n_3-t_30_200}, which is another time scale of Fig. \ref{x-eps_1-m_2-n_3-t_-2_4}, some fluctuations in spacetime's radius could exist before it becomes collapsed. It may pretend as an oscillating behavior before being crunched.

\subsection{Late Time Acceleration}
\begin{figure}
\centering
\includegraphics{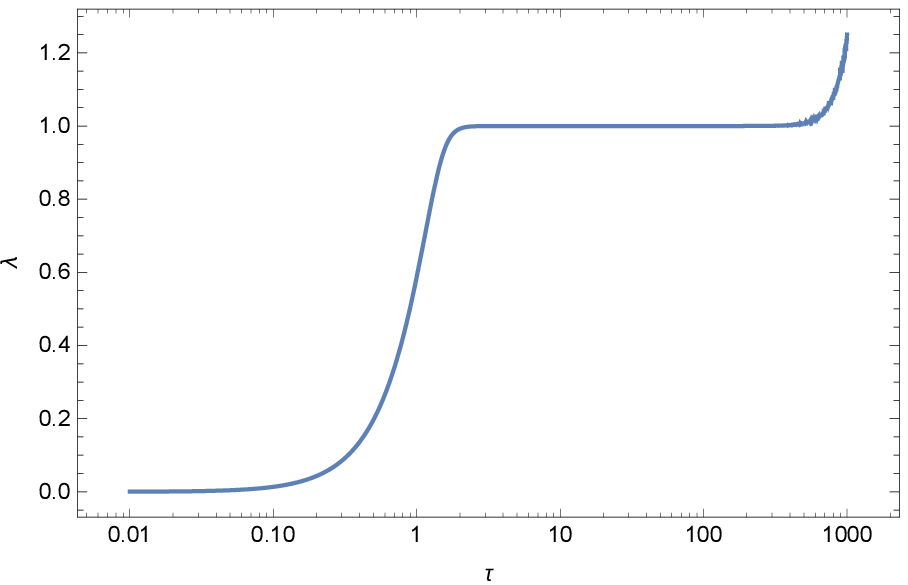}
\caption{\label{x-eps_1-m_3-n_4-t_-2_3} Evolution of horizon's radius of an accelerating model-based universe after experiencing a big bang and inflationary epochs. In this model, $m=3$, and $n=4$.}
\end{figure}

\begin{figure}
\centering
\includegraphics{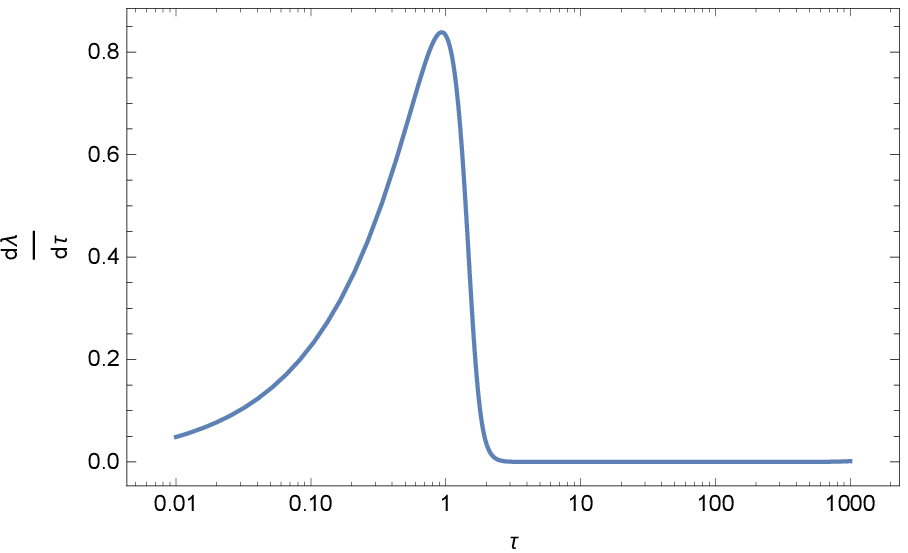}
\caption{\label{v-eps_1-m_3-n_4-t_-2_3} Evolution of horizon's velocity of an accelerating model-based universe after experiencing a big bang and inflationary epochs. In this model, $m=3$, and $n=4$.}
\end{figure}

\begin{figure}
\centering
\includegraphics{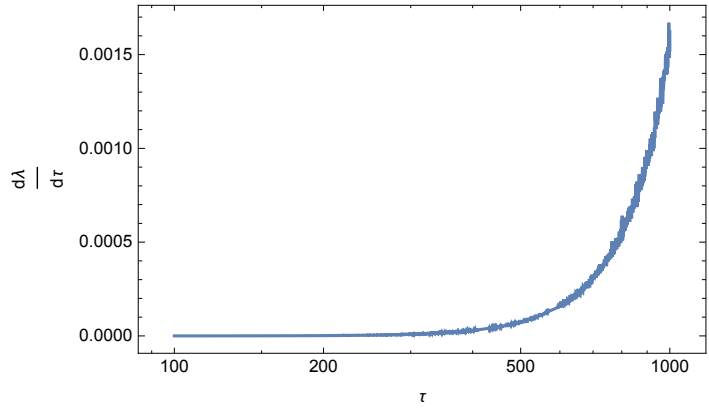}
\caption{\label{v-eps_1-m_3-n_4-t_2_3} A time scale of the horizon's velocity evolution of the model $m=3$ and $n=4$ illustrates a gradual increase in velocity.}
\end{figure}

The accelerating expansion of the universe is thought to have begun around four billion years ago, whenever the universe reached its dark energy-dominated phase. An accelerated expansion can be described by a particular value of the cosmological constant, analogous to positive vacuum energy, known as "dark energy" in general relativity. In de Sitter spacetime in the presence of a background velocity field, we study the case $m=3$ in Eq. (\ref{TM1}), which has six solutions of $g_{tt}=0$ from Eq. (\ref{dS2xtv}) with only one physical solution as
\begin{equation}\label{LTA}
\lambda(\tau)=\sqrt[3]{\frac{1}{2}} \Bigg(\frac{\sqrt[3]{\frac{2}{3}} \tau ^{2n+2} \left(\frac{9 n^4 \epsilon ^2}{\tau ^{4n+4}}+\sqrt{\frac{3 n^6 \epsilon ^3}{\tau ^{8n+8}} \left(27 n^2 \epsilon +4 \tau ^{2n+2}\right)}\right)^{2/3}-2 n^2 \epsilon }{n^2 \epsilon  \left(\frac{9 n^4 \epsilon ^2}{\tau ^{4n+4}}+\sqrt{\frac{3 n^6 \epsilon ^3}{\tau ^{8n+8}} \left(27 n^2 \epsilon +4 \tau ^{2n+2}\right)}\right)^{1/3}}\Bigg)^{1/2}.
\end{equation}
In this case, the horizon's expansion has a different behavior concerning the case $m=1$. As illustrated in Fig. \ref{x-eps_1-m_3-n_4-t_-2_3}, after a while, when inflation ended, another accelerating expansion happened. This case is similar to the late time accelerating expansion of the universe without considering dark energy. Horizon's velocity has shown in Fig. \ref{v-eps_1-m_3-n_4-t_-2_3} and Fig. \ref{v-eps_1-m_3-n_4-t_2_3} over different time scales. As it is unavailable to view late time velocity growing in Fig. \ref{v-eps_1-m_3-n_4-t_-2_3}, Fig. \ref{v-eps_1-m_3-n_4-t_2_3} depicts a gradual increase in expansion velocity of the model-depended universe.

\clearpage

\section{Conclusion}\label{C}

This article is started with the concept that velocity may be one of the fundamental foundations for the universe's expansion in a form of a new metric. In this case, we introduced it separately into the metric and investigated its consequences in Minkowski spacetime and de Sitter spacetime. The presence of a velocity field is not a novel assumption in discussions of the universe's expansion. Since the discovery of Hubble expansion, when it became evident that two points in the large-scale structure were moving apart, the presence of a rate of expansion of the hypothetical backdrop seemed unavoidable. However, this velocity field was thought to be the consequence of the Big Bang rather than its cause.

Nevertheless, in this work, it has been demonstrated that the existence of a velocity field from the beginning of time may provide consistent findings with observations and ideas about the universe. Ideas include justifying the theory of inflation, which demands limitless expansion in a limited period, and the concept of the universe's primordial black holes, which might be an alternative to the initial dark matter. Furthermore, despite a beginning velocity, the eventual fate of the universe, which in conventional cosmological models may accelerate expansion or catastrophic collapse, is foreseeable.

Of course, different types of velocity functions might have different results, according to calculations presented in this article. A generic velocity function proportional to $x/t^n$, where $n=1, 4, 12$, can induce a continuous expansion in a Minkowski spacetime and an inflationary epoch in a de Sitter background. In contrast, another velocity function proportional to $x^2/t^n$, where $n=3$, can cause the universe to collapse after experiencing an inflationary period. Both of these models also can produce primordial black holes. The divergence of the scalar curvature theoretically causes these types of black holes. Moreover, the existence of a universe with a late time accelerating expansion is equivalent to a velocity function proportional to $x^3/t^n$, where $n=4$. Then, it is possible to find a more general function that encompasses all of the above results. However, we did not seek a precise velocity function to justify all observed results in this research. Although it is exciting to discover a velocity function that supports all empirical data, the essential question is how to explain the existence of this velocity field in the context of general relativity or its alternative theories, which is a continuing problem. The world we study on the most significant possible scale, called cosmology, requires theories that justify massive inflation in a brief period and, on the other hand, have the content of about 95 percent of matter and energy is invisible to justify both the formation of the structure of the universe and the observed accelerated expansion. However, models such as the one introduced in this paper as a new metric can meet the needs of the standard cosmological model in the form of metric sentences in a geometric form. Given that the basis of general relativity is geometry, this type of research helps justify cosmological observations with the help of geometry, which will lead to a better understanding of general relativity and the prediction of extended models. The authors are investigating a Lagrangian-based method that has a velocity term as an outcome in metric sentences. In this topic, the additional terms in Einstein field equations will belong to the generalized Einstein-Hilbert action, which is under consideration.\\

\newpage
\clearpage

\newpage
\clearpage

\newpage

\clearpage

\newpage

\clearpage

\newpage

\clearpage

\newpage

\clearpage

\newpage

\clearpage

\newpage

\section*{Acknowledgments}

\bibliographystyle{amsplain}

\end{document}